\documentclass[mathleft]{an}
\usepackage{graphicx}
\usepackage{color, soul, ulem}
\usepackage{times}
\usepackage{natbib}

\begin{document}
\Pagespan{1055 }{}\Yearpublication{2011}\Yearsubmission{2011}\Month{}\Volume{332}\Issue{9/10} 
 \DOI{10.1002/asna.201111622}%

\title{Shock formation around planets orbiting M-dwarf stars}
\author{A. A. Vidotto\thanks{Corresponding author:\email{Aline.Vidotto@st-andrews.ac.uk}}, J.~Llama, M.~Jardine, Ch.~Helling \and K.~Wood}
\institute{SUPA, School of Physics and Astronomy, University of St Andrews, North Haugh, St Andrews, KY16 9SS, UK}

\titlerunning{Shock formation around planets orbiting M-dwarf stars}
\authorrunning{A. A. Vidotto et al.}

\received{}
\accepted{}
\publonline{later}


\abstract{Bow shocks can be formed around planets due to their interaction with the coronal medium of the host stars. The net velocity of the particles impacting on the planet determines the orientation of the shock. At the Earth's orbit, the (mainly radial) solar wind is primarily responsible for the formation of a shock facing towards the Sun. However, for close-in planets that possess high Keplerian velocities and are frequently located at regions where the host star's wind is still accelerating, a shock may develop ahead of the planet. If the compressed material is able to absorb stellar radiation, then the signature of bow shocks may be observed during transits. Bow-shock models have been investigated in a series of papers \citep{2010ApJ...722L.168V, 2011MNRAS.411L..46V, 2011MNRAS.414.1573V, 2011MNRAS.416L..41L} for known transiting systems. Once the signature of a bow-shock is observed, one can infer the magnetic field intensity of the transiting planet. Here, we investigate the potential to use this model to detect magnetic fields of (hypothetical) planets orbiting inside the habitable zone of M-dwarf stars. For these cases, we show, by means of radiative transfer simulations, that the detection of bow-shocks of planets surrounding M-dwarf stars may be more difficult than for the case of close-in giant planets orbiting solar-type stars.}

\maketitle

\section{Introduction}\label{sec.intro}
The interaction of a planet with the corona of its host star can give rise to the formation of shocks that surround the planet's magnetosphere. Similar to what occurs around the Earth and other planets in the solar system, bow-shocks may develop around exoplanets. This idea had recently been applied to explain the lightcurve asymmetry observed in the near-UV transit of the close-in giant planet WASP-12b.

\citet{2010ApJ...714L.222F} observed that the near-UV transit lightcurve of WASP-12b shows an early ingress when compared to its optical transit. Such observations indicate the presence of asymmetries in the exosphere of the planet. In particular, \citet{2010ApJ...722L.168V} suggested that this asymmetry could be explained by the presence of a shock surrounding the planet's magnetosphere. 

To test this idea, \citet{2011MNRAS.416L..41L} performed Monte Carlo radiation transfer simulations of the near-UV transit of WASP-12b. Their results support the hypothesis proposed by \citet{2010ApJ...722L.168V} as it explains both the observed level of absorption and the time of the (early) ingress observed in the near-UV lightcurve of the planet. 

An interesting outcome of this model is that one can probe the presence of planetary magnetic fields. Magnetic fields may provide an import shield against erosion caused by the impacting particles from the stellar wind. In the Earth, the presence of a magnetic field is associated with the development of life. M-dwarf stars are currently the main targets in searches for terrestrial habitable planets. Therefore, it is interesting to analyse whether shocks could be used to trace magnetic field of Earth-like planets orbiting M-dwarf stars.

The present paper is organised as follows. Section~\ref{sec.shock} summarizes the general characteristic of the bow-shock model that was initially developed for the case of WASP-12b. We also provide an overview of the results obtained through radiative transfer simulations, and discuss how this model could be useful to probe for planetary magnetic fields in other extra-solar planets. In Section~\ref{sec.mdwarf}, we investigate the potential for detecting bow shocks surrounding the magnetospheres of hypothetical transiting planets orbiting inside the habitable zones of M-dwarf stars.  Section~\ref{sec.conclusion} presents our conclusions.

\section{The Shock Model} \label{sec.shock}
\begin{figure}[h]
\includegraphics[width=80mm]{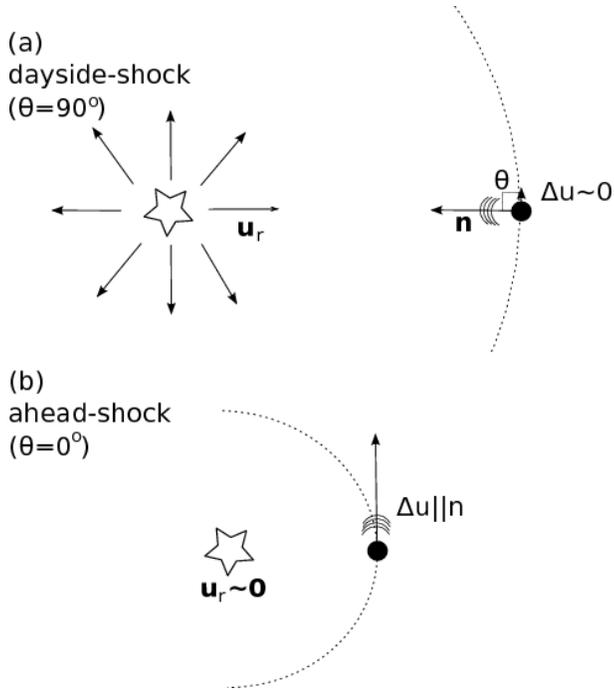}
\caption{Sketch of shock limits: (a) dayside-shock ($\theta=90^{\rm o}$), (b) ahead-shock ($\theta=0^{\rm o}$). Arrows radially leaving the star depict the stellar wind, dashed semi-circles represent the orbital path, $\theta$ is the deflection angle between the shock normal ${\bf n}$ and the relative azimuthal velocity of the planet $\Delta {\bf u}$. Adapted from \citet{2010ApJ...722L.168V}. \label{fig.shockconditions} }
\end{figure}

A bow shock around a planet is formed when the relative motion between the planet and the stellar corona/wind is supersonic. The shock configuration depends on the direction of the flux of particles that arrives at the planet. We illustrate two different limits of the shock configuration in Fig.~\ref{fig.shockconditions}, where $\theta$ is the deflection angle between the azimuthal direction of the planetary motion and ${\bf n}$ is a vector that defines the outward direction of the shock. As seen from the planet, $-{\bf n}$ is the velocity of the impacting material. 

The first shock limit, a ``dayside-shock'', occurs when the dominant flux of particles impacting on the planet arises from the (mainly radial) wind of its host star. For instance, the impact of the supersonic solar wind forms a bow shock at the dayside of Earth's magnetosphere (i.e., at the side that faces the Sun). This condition is illustrated in Fig.~\ref{fig.shockconditions}a and is met when 
\begin{equation}
u_r > c_s,
\end{equation}
 where $u_r$ and $c_s$ are the local radial stellar wind velocity and sound speed, respectively.

A second shock limit, an ``ahead-shock'' (Fig.~\ref{fig.shockconditions}b), occurs when the dominant flux of particles impacting on the planet arises from the relative azimuthal velocity between the planetary orbital motion and the ambient plasma. This condition is especially important when the planet orbits at a close distance to the star, and therefore, possesses a high Keplerian velocity $u_K$. In this case, the velocity of the particles that the planet ``sees'' is supersonic if 
\begin{equation}
\Delta u = |u_K-u_\varphi|> c_s,
\end{equation}
where $u_\varphi$ is the azimuthal velocity of the stellar corona.

{ As we show later on Section 3, planets orbiting in the habitable zones of M-dwarf stars fall into an intermediate case, where both the wind and the azimuthal relative velocities will contribute to the formation of a shock around the planet. }

\subsection{Near-UV Early Ingress of WASP-12b}
Motivated by observations made with the HST \citep{2010ApJ...714L.222F}, which showed that in the near-UV, the transit of WASP-12b starts earlier than in the optical wavelengths, we applied the model presented in Section~\ref{sec.shock} to the particular case of WASP-12b \citep{2010ApJ...722L.168V}. 

\begin{figure}
	\includegraphics[width=80mm]{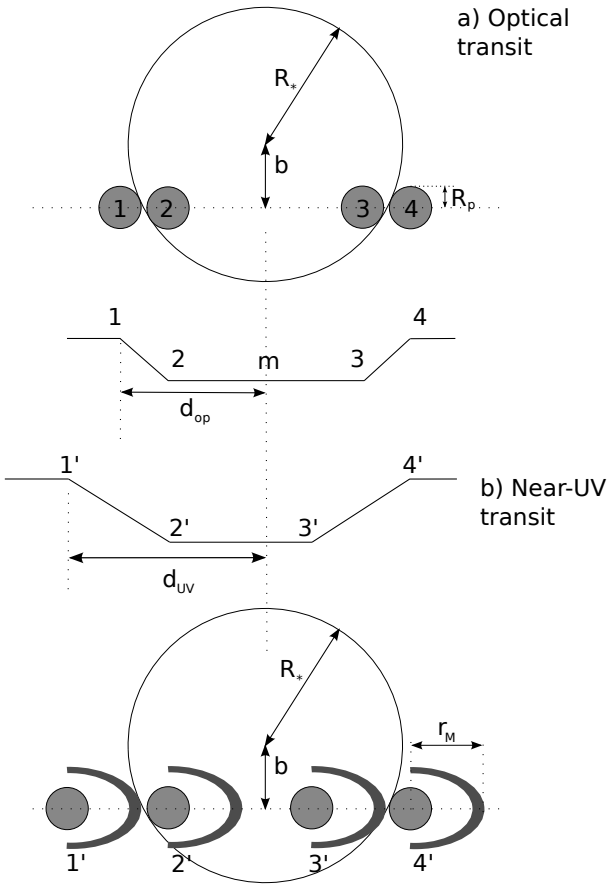}
\caption{Sketches of the light curves obtained through observations in the a) optical and b) near-UV, where the bow shock surrounding the planet's magnetosphere is also able to absorb stellar radiation. Figure adapted from \citet{2011MNRAS.414.1573V}.}\label{fig.sketch2}
\end{figure}

WASP-12b orbits a late-F main-sequence star with mass $M_* = 1.35~M_\odot$ and radius $R_*=1.57~R_\odot$, at an orbital radius of $a=3.15~R_*$ \citep{2009ApJ...693.1920H}. Due to its close proximity to the star, the flux of coronal particles impacting on the planet should come mainly from the azimuthal direction, as the planet moves at a Keplerian orbital velocity of $u_K = (G M_*/a)^{1/2}\sim 230$~km~s$^{-1}$ around the star. Therefore, stellar coronal material is compressed ahead of the planetary orbital motion, possibly forming a bow shock ahead of the planet. \citet{2010ApJ...722L.168V}'s suggestion is that this material is able to absorb  enough stellar radiation, causing the early-ingress observed in the near-UV light curve (see Figure~\ref{fig.sketch2}).
 
By measuring the phases at which the near-UV and the optical transits begin, one can derived the stand-off distance from the shock to the centre of the planet. In the geometrical consideration made next, we assume that the planet is fully superimposed on the disk of the central star, which is a good approximation for, e.g., small planets and transits with small impact parameters. Consider the sketches presented in Fig.~\ref{fig.sketch2}, where $d_{\rm op}$ and $d_{\rm UV}$ are, respectively, the sky-projected distances that the planet (optical) and the system planet+magnetosphere (near-UV) travel from the beginning of the transit until the middle of the optical transit
\begin{equation}\label{eq.dop}
d_{\rm op} = (R_*^2 - b^2)^{1/2} + R_p
\end{equation}
and
\begin{equation}\label{eq.duv}
d_{\rm UV} = (R_*^2 - b^2)^{1/2} + r_M,
\end{equation}
where $b$ is the impact parameter derived from transit observations, $R_p$ is the planetary radius, and $r_M$ is the distance from the shock nose to the center of the planet. The start of the optical transit occurs at phase $\phi_1$ (point $1$ in Fig.~\ref{fig.sketch2}a), while the near-UV transit starts at phase $\phi_{1'}$ (point $1'$ in Fig.~\ref{fig.sketch2}b). Taking the mid-transit phase at $\phi=\phi_m \equiv 1$, we note that $d_{\rm op}$ is proportional to $(1-\phi_1)$, while $d_{\rm UV}$ is proportional to $(1-\phi_{1'})$. Using Equations~(\ref{eq.dop}) and (\ref{eq.duv}), the stand-off distance $r_M$ is derived from observed quantities
\begin{equation}\label{eq.rm}
r_M =  \frac{1-\phi_1}{1-\phi_{1'}}[(R_*^2 - b^2)^{1/2} + R_p] - (R_*^2 - b^2)^{1/2}. 
\end{equation}

We assume the stand-off distance to trace the extent of the planetary magnetosphere. At the magnetopause, pressure balance between the coronal total pressure and the planet total pressure requires that
\begin{equation}\label{eq.equilibrium}
{\rho_c \Delta u^2} + \frac{[B_c(a)]^2}{8\pi} + p_c= \frac{[B_{p}(r_M)]^2}{8\pi} + p_{p} ,
\end{equation}
where $\rho_c$, $p_c$ and $B_c(a)$ are the local coronal mass density, thermal pressure, and magnetic field intensity, and $p_p$ and $B_p (r_M)$ are the planet thermal pressure and magnetic field intensity at $r_M$. In the case of a magnetised planet, the planet total pressure is usually dominated by the contribution caused by the planetary magnetic field (i.e., $p_p \sim 0$). 

\citet{2010ApJ...722L.168V} showed that, because WASP-12b is at a close distance to the star, the kinetic term of the coronal plasma may be neglected in Eq.~(\ref{eq.equilibrium}). They also neglected the thermal pressure, so that Eq.~(\ref{eq.equilibrium}) reduces to $B_c(a) \simeq B_p(r_M)$. Further assuming that stellar and planetary magnetic fields are dipolar, we have
\begin{equation}\label{eq.bplanet}
B_p = B_* \left(  \frac{R_*/a}{R_{p}/r_M}  \right)^3 ,
\end{equation}
where $B_*$ and $B_p$ are the magnetic field intensities at the stellar and planetary surfaces, respectively. 

Therefore, by determining the phase at which the near-UV transit begins, one can derive the stand-off distance (Eq.~\ref{eq.rm}) and then estimate the intensity of the magnetic field of the planet (Eq.~\ref{eq.bplanet}), provided that the stellar magnetic field is known. 

For WASP-12, we use the upper limit of $B_*<10$~G \citep{2010ApJ...720..872F} and the stand-off distance obtained from the near-UV transit observation $r_M=4.2~R_p$ \citep{2010ApJ...721..923L}, and we predict a planetary magnetic field of $B_p < 24$~G for WASP-12b.

\subsection{Radiation Transfer Simulations of the near-UV transit}
In order to test the hypothesis of the bow-shock model was indeed able to cause the lightcurve asymmetry observed in WASP-12b, \citet{2011MNRAS.416L..41L} performed Monte Carlo radiation transfer simulations of the near-UV transit of WASP-12b. The characteristics of the stellar coronal plasma (density, velocity and temperature), modelled by \citet{2010ApJ...722L.168V}, were adopted in order to derive the density at the shock nose and the angle at which the shock is formed. As in \citet{2010ApJ...722L.168V}, \citet{2011MNRAS.416L..41L} assumed a shock in the adiabatic limit where the density behind such a shock increases by a factor of four with respect to the density ahead of the shock (stellar coronal material). 

The characteristics of the local plasma surrounding the planet were derived based on simple models that we describe next. \citet{2010ApJ...722L.168V} adopted two scenarios. The first one considers the corona as a hydrostatic medium, so that it corotates with the star. The other one, considers that the corona is filled with an expanding, isothermal wind. In this case, the wind radial velocity $u_r$ is derived from the integration of the differential equation along the radial coordinate $r$
\begin{equation}  \label{eq.windsolution}
u_r \frac{\partial u_r}{\partial r} = -\frac{1}{\rho_c} \frac{\partial p_c}{\partial r} - \frac{G M_*}{r^2} , 
\end{equation} 
where $G$ is the gravitational constant. The first prescription adopted has the advantage of having analytical solutions. The latter one, although lacking an analytical expression, can be easily integrated. In the first scenario, there is no radial velocity of the wind plasma, so that the shock forms ahead of the planet, while in the second scenario, the shock forms at an intermediate angle. These angles were  used in the simulations of \citet{2011MNRAS.416L..41L}.

To compute the near-UV lightcurve, the characterisation of the three-dimensional geometry of the shock is required. Therefore, two unknowns of the shock geometry, its solid angle and its thickness, had to be specified. In order to tackle the influence of these two parameters, \citet{2011MNRAS.416L..41L} performed simulations for several shock geometries. As a result, they found that different sets of parameters could produce similar solutions.

To constrain the model parameter, \citet{2011MNRAS.416L..41L} relied on the information present in the near-UV lightcurve by \citet{2010ApJ...714L.222F}. By analysing different models that were equally able to provide a good fit to the HST data, they were able to place constraints on 1) the phase of the near-UV ingress ($\phi_{1'}$)  and 2) the optical depth of the shocked material, related to the extent of the shock and its thickness. From 1), they could constrain the projected stand-off distance as $5.5~R_p$, slightly larger than the value derived by \citet{2010ApJ...721..923L} and used in \citet{2010ApJ...722L.168V}. From 2), they showed that the shocked material does not need to have a large optical depth to cause the amount of absorption observed in the near-UV HST lightcurve.  

The simulations presented by \citet{2011MNRAS.416L..41L} support the hypothesis that a bow-shock could generate an early ingress of the transit, as the addition of a bow shock breaks the symmetry of the transit lightcurve. Nevertheless, the current data is not yet adequate to fully test this prediction. Near-UV observations of WASP-12b (and other exoplanets, see next are desirable in order to test and constrain the models.

\subsection{Magnetic Fields in Other Exoplanets} \label{sec.othersystems}
To extend the previous model to other transiting systems, near-UV data must be acquired. \citet{2011MNRAS.411L..46V} presented a classification of the known transiting systems according to their potential for producing shocks that could cause observable light curve asymmetries. The data used are from the compilation in http://exoplanet.eu (Sept/2010) and the sky-projected stellar rotation velocities $v_{\rm rot} \sin(i)$ from \citet{2010ApJ...719..602S}, assuming that $\sin(i)\simeq 1$. 

\begin{figure}[h]
\includegraphics[width=80mm]{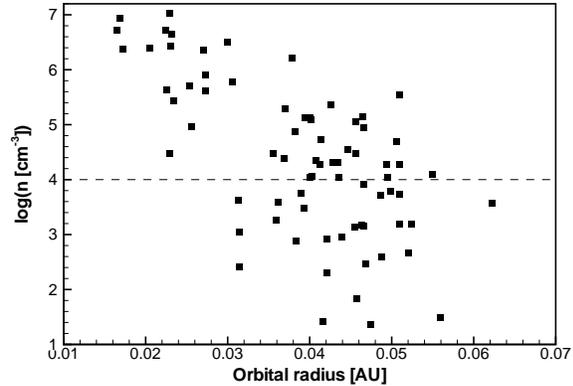}
\caption{Predicted coronal density at the orbit of each transit planet known as of Sept/2010, assuming that the stellar magnetic field is strong enough to confine the coronal gas out to the orbit of the planet. 
The dashed line represents a lower limit for detection of bow shocks. Figure adapted from \citet{2011MNRAS.411L..46V}}\label{dens_all}
\end{figure}

Once the conditions for shock formation are met, to be detected, it must compress the local plasma to a density high enough to cause an observable level of optical depth. For a hydrostatic, isothermal corona, the local density at the orbital radius $a$ is
\begin{eqnarray}\label{eq.dens-hyd}
 \frac{n}{n_0} = \exp \left\{ \frac{u_K^2}{c_s^2}\left[ 1-\frac{a}{R_*}\right] + \frac{v_{\rm rot}^2}{c_s^2}\left[\frac{a}{R_*}-1\right] \right\},
\end{eqnarray}
where $n_0$ is the number density at the base of the corona. The maximum coronal temperature that could still allow shock formation is such that $c_{s} > |u_{\varphi} - u_K|$, i.e., 
\begin{equation}\label{eq.tmax}
T_{\rm max} = {( u_{\varphi} - u_K)^2 m}/{k_B},
\end{equation}
where the isothermal sound speed is $c_s = k_B T/m$, $k_B$ is the Boltzmann constant, and $m$ the particle mass. If we assume that the corona corotates with the star, at the position of the orbital radius of a close-in planet, $u_{\varphi} = v_{\rm rot}{a}/{R_*}$. By adopting the maximum temperature that could still allow shock formation, we can obtain a minimum density required for shock formation using Equation (\ref{eq.dens-hyd}). Fig.~\ref{dens_all} shows this critical density as a function of orbital radius for a range of planets. This result was produced assuming that all stars have a base coronal density equal to that of the Sun ($10^8~{\rm cm}^{-3}$), which is a reasonable approximation for stars with spectral types similar to that of our Sun. The most promising candidates to present shocks are: WASP-19b, WASP-4b, WASP-18b, CoRoT-7b,  HAT-P-7b, CoRoT-1b, TrES-3, and WASP-5b.

\citet{2011MNRAS.411L..46V} estimated that a lower limit of the stellar coronal density $n_{\rm min} \simeq 10^4$~cm$^{-3}$  could still provide detection (dashed line). Taking that into account, we note that a reasonable number of planets should lie above this density-detection threshold, suggesting that a detectable shock might be a common feature surrounding transiting planets.

\section{Shock Formation around Planets Orbiting M-dwarf Stars} \label{sec.mdwarf}
Current and future missions like Kepler, CoRoT and Plato are designed to search for Earth-like planets. Low-mass stars ($\sim 0.06$--$0.8M_\odot$) are currently the main targets in searches for terrestrial habitable planets as their low brightness provide good contrast for detection of smaller-radius planets (transit method) and their low masses facilitate detection of lower-mass planets (radial velocity method). These methods favour detection of planets orbiting close to their host star, which in the case of low-mass stars is also the region where the habitable zone (HZ) is located. The concept of habitable zones relies on the location where liquid water may be found, which is essentially determined by the luminosity of the star. Because M-dwarf stars are low-luminosity objects, the habitable zone is believed to lie near the star. The habitable zone for a $0.5M_\odot$ star is believed to lie between $0.2$ -- $0.5$~AU, while for a $0.1M_\odot$ star, it is thought to lie much closer, between $0.02$ -- $0.05$~AU \citep{1993Icar..101..108K}.

However, even if a planet is in the HZ of a low-mass star, the interaction of the planet with the host star's wind may erode its atmosphere, removing an important shield for the creation and development of life. The interaction of a planet with its host star wind can produce a shock surrounding the planet. As presented before, if we are able to detect the stand-off distance between the shock and a planet (e.g., as during near-UV transit observations), we are then able to estimate the planet's magnetic field intensity. 

\begin{figure*}
\mbox{
\includegraphics[width=80mm]{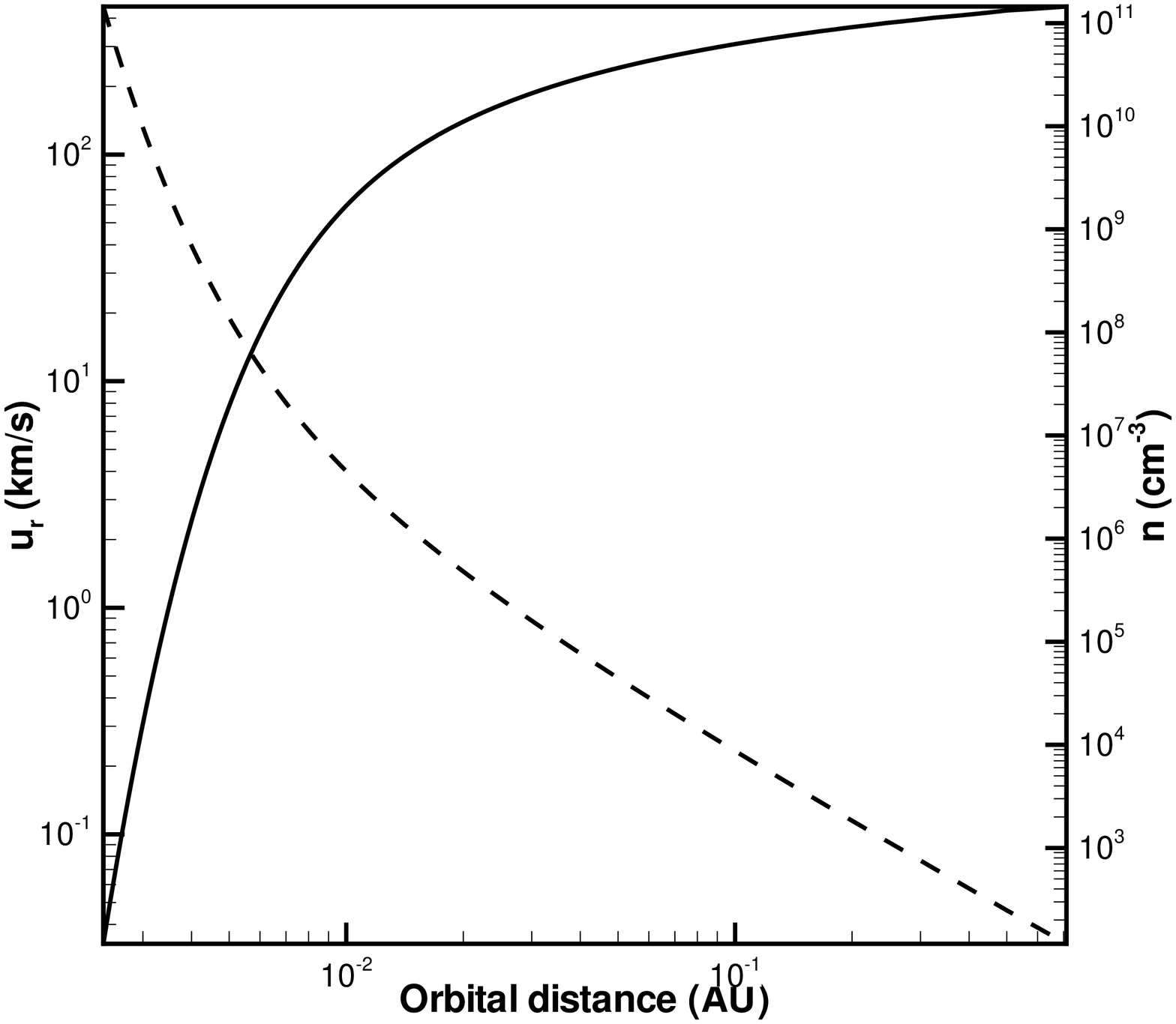} 
\includegraphics[width=80mm]{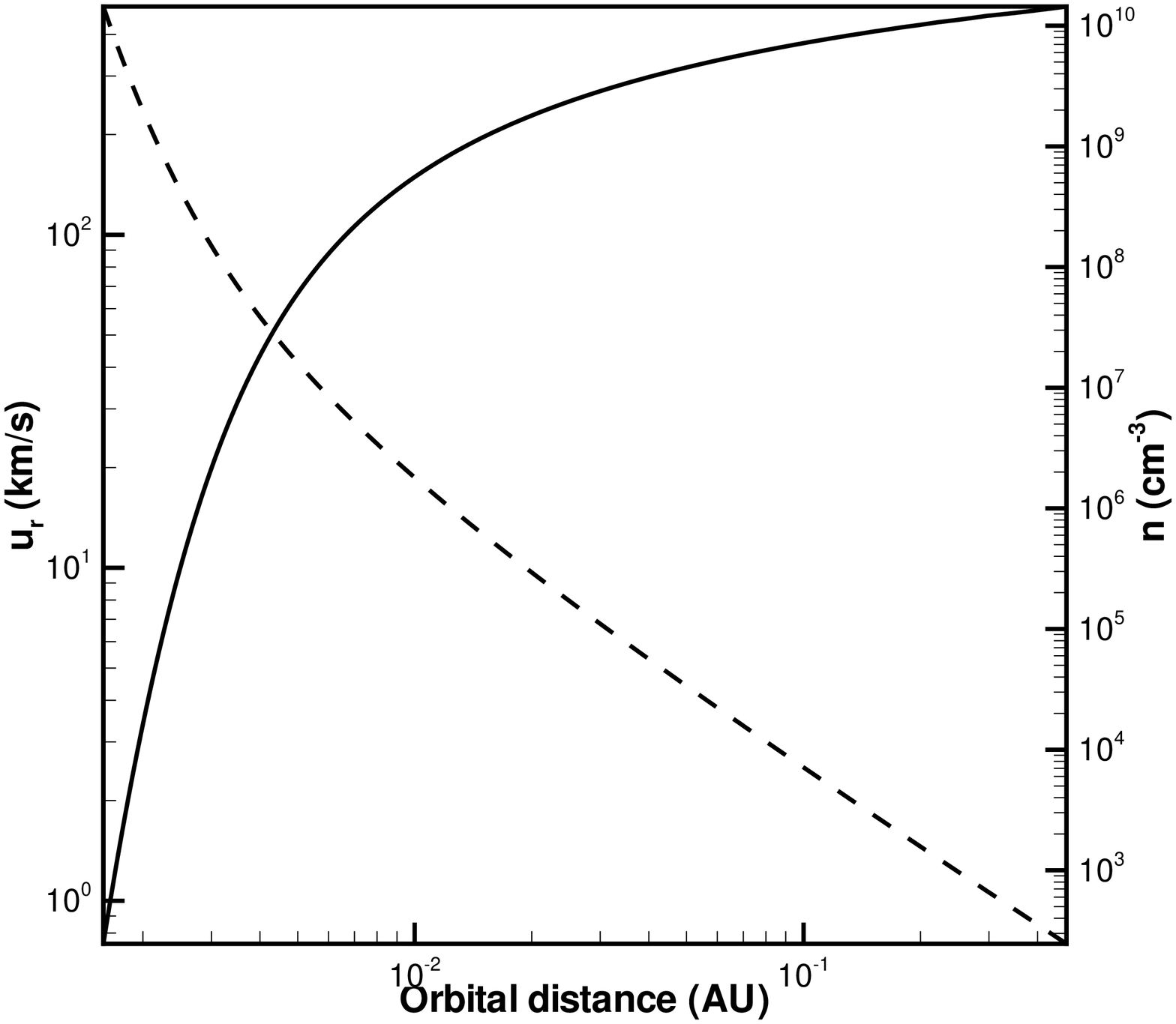}}
\mbox{\includegraphics[width=80mm]{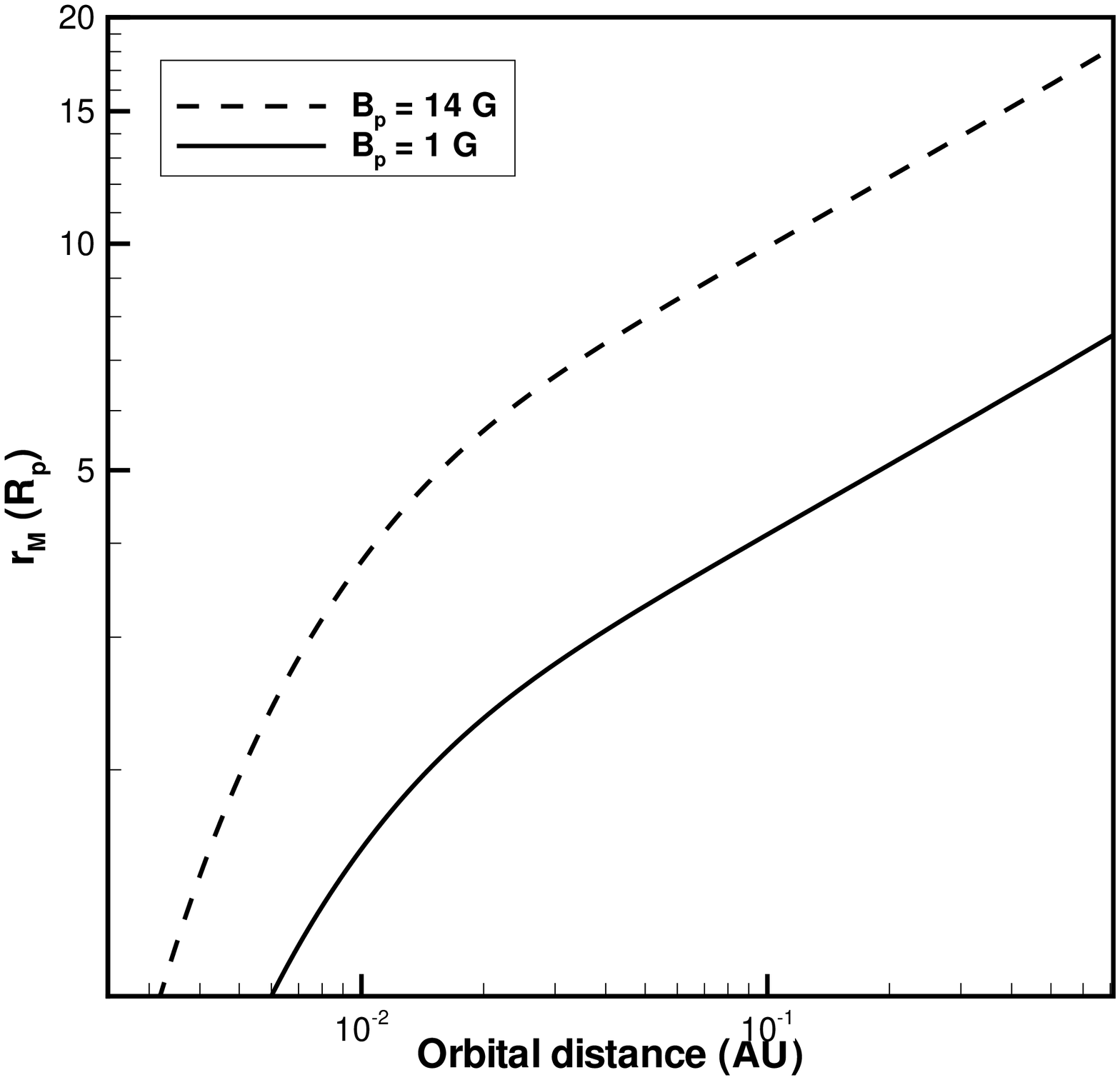} 
\includegraphics[width=80mm]{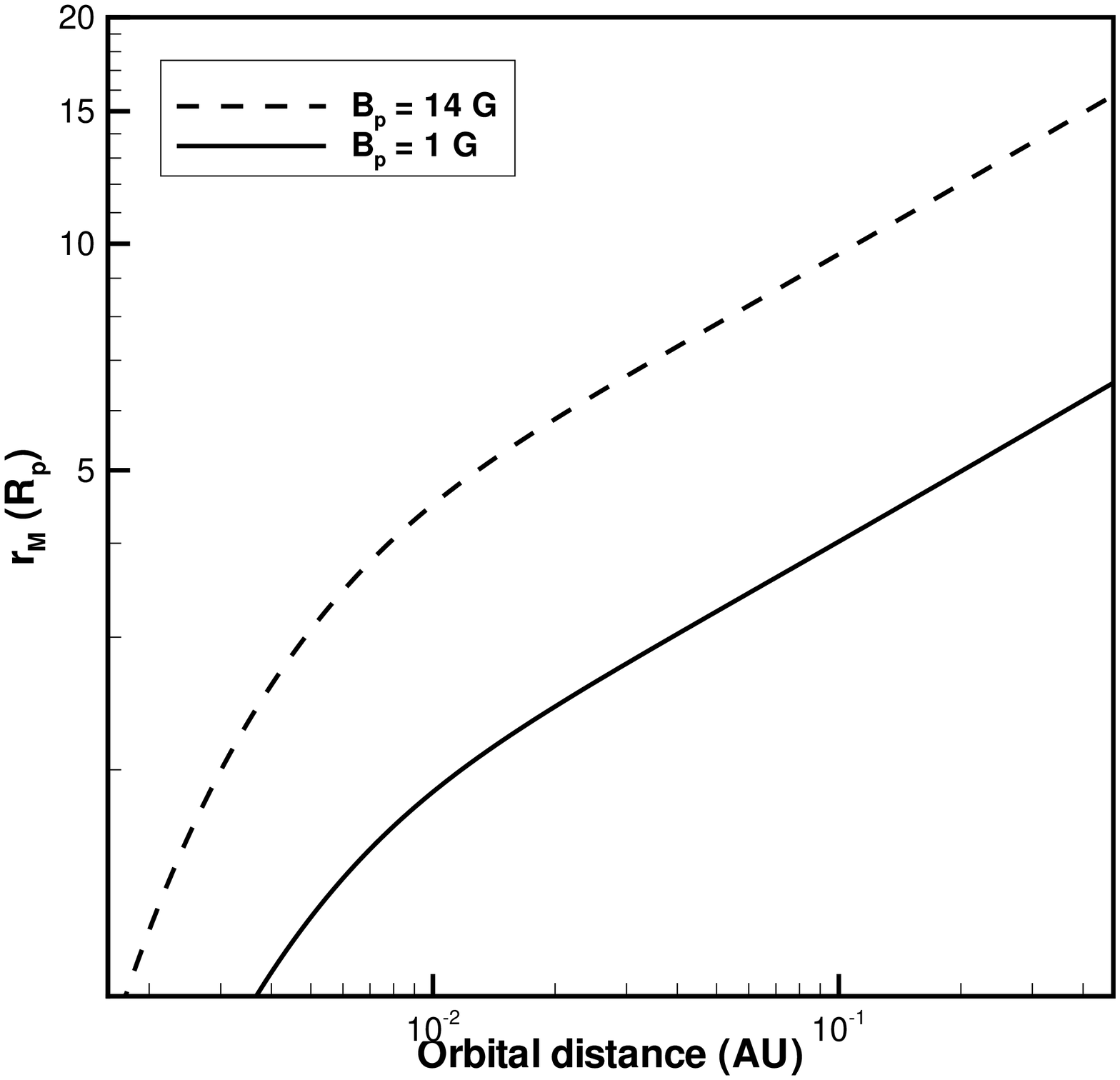}}
\mbox{\includegraphics[width=80mm]{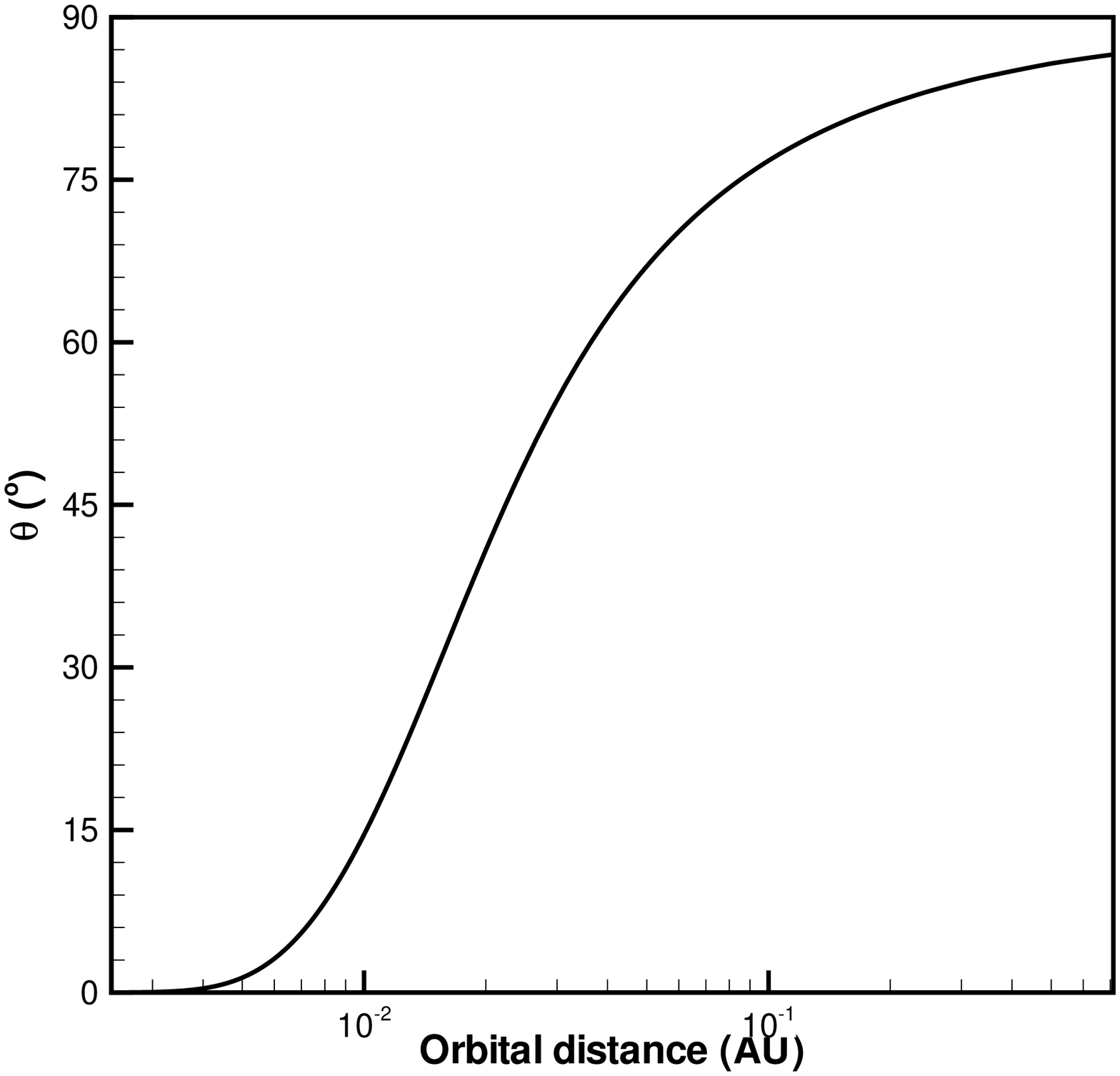} 
\includegraphics[width=80mm]{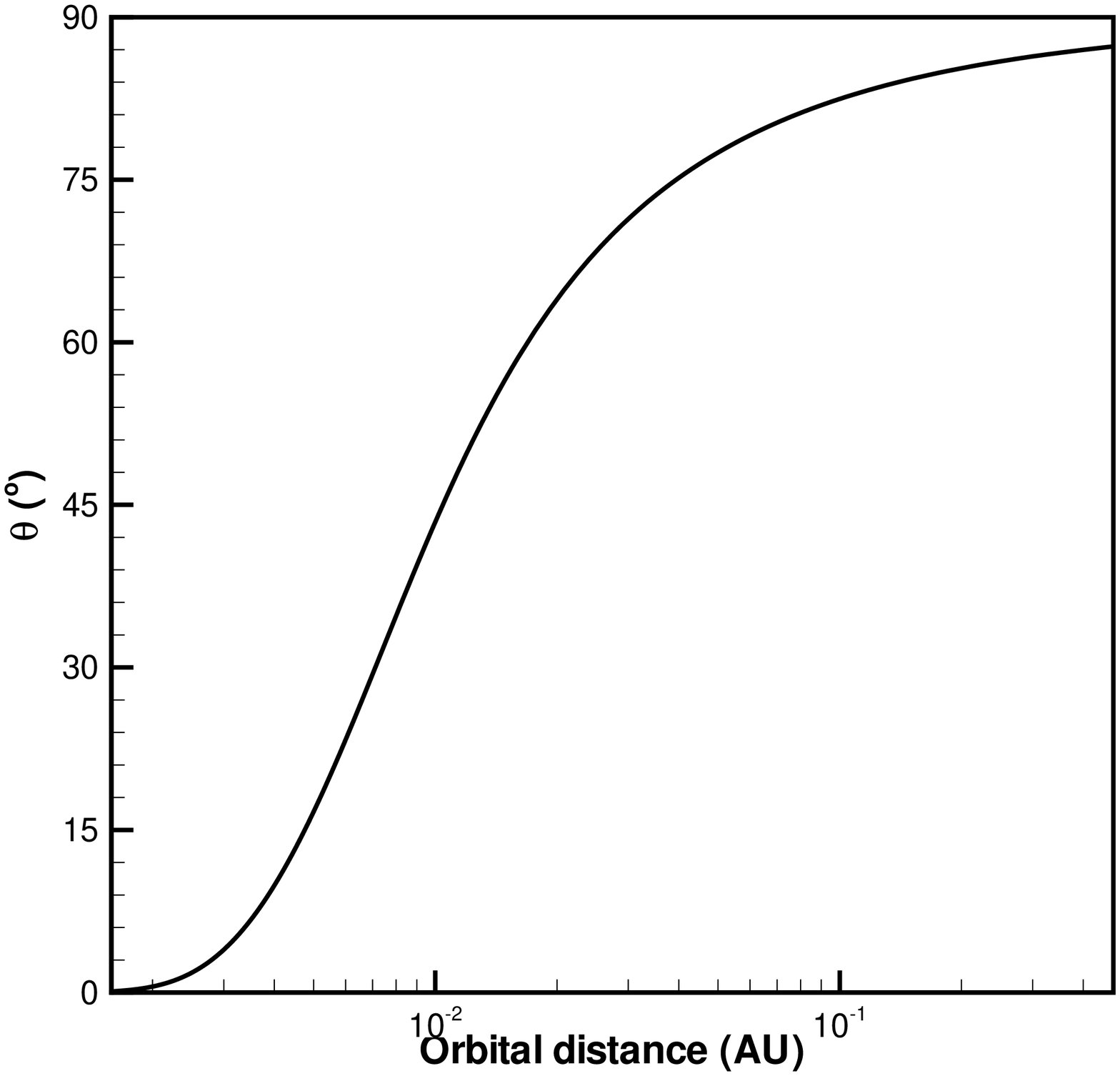}
}
\caption{Results obtained from our wind models for the M1.5 case (left) and M4.0 case (right). Top panels: wind velocity solution (solid line) and the density profile (dashed line). Middle panels: magnetospheric size as a function of orbital radius assuming a planet with a magnetic field similar to Jupiter ($B_p=14$~G, dashed line) and with a magnetic field similar to Earth ($B_p=1$~G, solid line). Bottom panels: the shock angle as a function of the orbital radius of a planet. }\label{resultsdM}
\end{figure*}

In this section, we calculate the characteristics of the wind of two M-dwarf stars with spectral types M1.5 and M4.0, adopting a thermally driven wind \citep[Equation~(\ref{eq.windsolution}), ][]{1958ApJ...128..664P}. We assume these winds are isothermal with temperature $T=1\times 10^6~$K and their densities are such that their mass-loss rates are $\dot{M}=10^{-13}~\rm{M}_\odot ~\rm{yr}^{-1}$. Table~\ref{table} shows the characteristics adopted for the star and the planet in our analysis. The top panels in Figure~\ref{resultsdM} show the wind velocity solution (solid line) and the density profile (dashed line) for the M1.5 case (left) and the M4.0 case (right). 

\begin{table}
\caption{Values adopted in our models (upper part) and results obtained (lower part).
}
\label{tab.values}
\begin{tabular}{lcc}\hline
Quantity & M1.5 & M4.0\\ \hline
$M_\star (M_\odot)$ & $0.57$ & $0.28$\\
$R_\star (R_\odot)$ & $0.51$ & $0.34$\\
\hline
$a$ (AU) & $0.2$ & $0.03$\\
$\theta (^{\rm o})$ & $82$ & $71$ \\
$(r_M/R_p) {[B_{p}=14~{\rm G}]}$ & $12.2$ & $6.7$ \\
$(r_M/R_p) [B_p=1~{\rm G}]$ & $5.1$ & $2.8$ \\
$n_{\rm Mg, shock}$ (cm$^{-3}$) & $0.09$ & $5.6$ \\
\hline
\end{tabular}\label{table}
\end{table}

\subsection{Deriving the Magnetospheric Characteristics}
In the calculations presented next, we assume the planet to be in a circular orbit. We use Equation~(\ref{eq.equilibrium}) to derive the planetocentric distance to the magnetopause, such that
\begin{equation}\label{eq.rm-theor}
\frac{r_M}{R_p}=\left(\frac{B_p^2}{8\pi P_{\rm eff}} \right)^{1/6},
\end{equation}
where we prescribed the planetary magnetic field topology to be dipolar. The total pressure exerted in the planet due to the coronal wind $P_{\rm eff}$ is 
\begin{equation}\label{eq.peff}
P_{\rm eff} = \rho_c \Delta u^2 + \frac{[B_c(a)]^2}{8\pi} + p_c .
\end{equation}
Our wind solutions neglect the presence of the stellar magnetic field ($B_c =0$). We note that the local (i.e., at $r=a$) coronal wind velocity $u_r$ and density $\rho_c$ are obtained from the wind models. In the case of WASP-12b, data allowed us to estimate $r_M$ and then derive $B_p$. Here, we take the inverse procedure. 

The middle panels in Figure~\ref{resultsdM} shows the magnetospheric size the planets orbiting the M-dwarf stars analysed here would present as a function of orbital radius. We considered two different cases and assumed a planet with a magnetic field similar to Jupiter ($B_p=14$~G, dashed lines) and with a magnetic field similar to Earth ($B_p=1$~G, solid lines). For the M1.5 case, a planet orbiting at $a=0.2$~AU (in its HZ) would present a magnetospheric size of  $r_M =12.2 R_p$ and $5.1~R_p$ for $B_p=1$~G and $14$~G, respectively. For the M4.0 case, the same hypothetical planet orbiting at $a=0.03$~AU (in its HZ) would present $r_M=6.7~R_p$ or $2.8~R_p$ for $B_p=1$~G and $14$~G, respectively. 

The orientation of the shock normal depends not only on the wind velocity, but also on the Keplerian velocity of the planet and on the azimuthal velocity of the wind. In the Parker wind solution, rotation is neglected. So the cases we study here are valid for slowly rotating stars. This angle is given by
\begin{equation}  \label{eq.theta}
\theta = {\rm atan}{ \left( \frac{u_r}{u_K} \right)} 
\end{equation} 
and is $\theta =90^{\rm o}$ for a dayside-shock and $\theta =0^{\rm o}$ for an ahead-shock (Figure~\ref{fig.shockconditions}). The bottom panels in figure~\ref{resultsdM} shows the shock angle as a function of the orbital radius of a planet. As we can see, the farther out the planet is, the shock tends to approach the dayside-type shock. For a planet at $a=0.2$~AU orbiting the M1.5 case, this angle is $\theta=82^{\rm o}$, while for  a planet at $a=0.2$~AU orbiting the M4.0 case, this angle is $\theta=71^{\rm o}$. We note that, due to the geometry of the transiting systems, dayside-shocks are more difficult to be detected as, during transit, the planet occults the shock from the observer. Furthermore, this type of shock may not be able to brake the symmetry of the transit lightcurve.  Nevertheless, for the hypothetical planet orbiting inside the HZ of the M4.0 star, the angle we found ($\theta=71^{\rm o}$) could still produce an early ingress.

\subsection{Detecting Bow-Shocks}
As discussed in Section~\ref{sec.othersystems}, the density of the shocked material should be high enough as to produce a significant level of absorption for it to be detected. Using the density derived in our wind models, we calculate the density in magnesium assuming the stars have solar metallicity.\footnote{We calculate the density of magnesium as a proxy for the detection of the shock in the near-UV, in analogy to the near-UV transit observed for WASP-12b \citep{2010ApJ...714L.222F}.} Because the shock compresses material ahead of the planetary motion, the density of the shocked material increases by a certain factor. For a strong shock in the adiabatic limit, this increase has its maximum value of $4$. Using the results for the wind models we derived, we estimate this increase to be of a factor of $3$ for both stars analysed here. Therefore, the densities of the shocked material for both cases are shown in Table~\ref{table}. As it can be seen, these values are considerably smaller than the value derived for WASP-12b $n_{\rm Mg}\sim 400$~cm$^{-3}$ \citep{2010ApJ...722L.168V}, indicating that the detection of bow-shocks of planets surrounding M-dwarf stars may be more difficult. 

Using radiative transfer simulations similar to the ones presented in \citet{2011MNRAS.416L..41L}, we computed the lightcurve of a planet orbiting in the HZ of the stars modeled here (values are shown in Table~\ref{table}). With reasonable assumptions for the shock extension and thickness, a bow-shock surrounding a $2~R_{\rm Earth}$-planet can not be detected, as it does not present a sufficient amount of optical depth that causes a detectable absorption. 

\begin{figure}[h]
\centering{
\includegraphics[width=83mm]{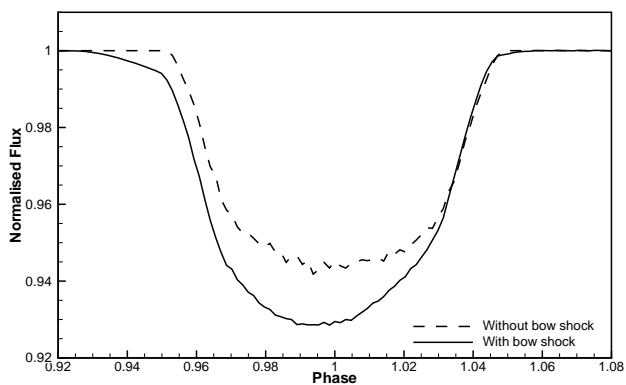}}
\caption{Lightcurves of a $2~R_{\rm Neptune}$-planet orbiting a M4.0-star at an orbital radius $a=0.03~$AU. Dashed line shows the lightcurve produced when no bow-shock surrounds the planet and the solid line is the result when the bow shock is included in the Monte Carlo radiation transfer simulation.}\label{fig.midtransit}
\end{figure}

It is worth noting that our results depend on the conditions adopted for the wind. For instance, mass-loss rates in M-dwarf stars are poorly constrained. A higher value of mass-loss rate would result in a higher value of the wind density, and therefore increased density in the shocked material that could be detected during transit. To illustrate that, we computed the lightcurve of a $2~R_{\rm Neptune}$-planet orbiting a M4.0-star at an orbital radius $a=0.03~$AU. We adopted a density for the wind (and for the shock) that is one order of magnitude larger than the one shown in Table~\ref{table}, implying that such a star would present a mass-loss rate of $\dot{M}=10^{-12}~\rm{M}_\odot ~\rm{yr}^{-1}$. We assumed that the planetocentric distance to the magnetopause is $r_M=5~R_p$ (or equivalently, $B_p \simeq 18$~G), the shock thickness is $1.5~R_p$ and its area lies within a solid angle of $\Delta \theta =50^{\rm o}$. The shock normal is oriented at $\theta = 71^{\rm o}$. Figure~\ref{fig.midtransit} shows a comparison of two synthetic lightcurves: one case where no bow shock is present (dashed line) and one case with bow shock (solid line). Note that our choice of parameters for the fictitious planet leads to a transit depth of $\sim 5$~percent normalised to the out-of-transit flux for the case without a bow shock and a $\sim 7$~percent-depth for the case where the bow shock was considered in the simulations. We  note that the early-ingress in this case is less-pronounced than in the case of WASP-12b \citep{2011MNRAS.416L..41L}.

We also note that M dwarf stars can exhibit substantial activity, which can lead to variations in the ambient medium surrounding the planet. In response to these variations, the planet's magnetosphere will adjust itself. Because of that, transit lightcurves in the near-UV can present variability \citep{2011MNRAS.414.1573V}.

\section{CONCLUSIONS}\label{sec.conclusion}

In this paper, we investigated the potential for detecting bow shocks surrounding the magnetospheres of transiting planets orbiting inside the habitable zones of M-dwarf stars. The idea that bow shocks may be present around exoplanets have now been developed in a series of papers \citep{2010ApJ...722L.168V,2011MNRAS.411L..46V,2011MNRAS.414.1573V,2011MNRAS.416L..41L}. Due to the interaction of the planet with the wind of its host star, bow shocks may form surrounding the magnetosphere of planets, and, if observable, the shock can help constrain planetary magnetic fields.

In order to analyse whether shock could be formed on planets orbiting M-dwarf stars, we applied \citet{2010ApJ...722L.168V}'s model to two hypothetical low-mass stars with spectral types M1.5 and M4.0. Earth-size planets were assumed to orbit inside the habitable zones of their host-stars and the stars were assumed to have a thermally driven-type wind. We showed that shock can be formed surrounding the planetary magnetosphere. For the case of a planet orbiting at $0.2$~AU from the M1.5 star, we showed that the shock forms at an angle closer to $\theta = 90^{\rm o}$, implying that the shock should be occulted from the observer during transit. If this is the case, the presence of the shock may not brake the optical lightcurve symmetry. For the case of a planet orbiting at $0.03$~AU from the M4.0 star, the shock forms at a smaller angle $\theta = 71^{\rm o}$. 

However, for the shock to be detected by means of an early-ingress, it has to compress the stellar coronal material to a level of optical depth that can cause enough absorption during transit. For the wind characteristics we assumed, the density we computed in the shocked material does not seem to be significantly high to cause an observable early-ingress in the planet near-UV lightcurve. This result was confirmed by means of radiative transfer simulations. This implies that the method applied to close-in giant planets orbiting solar-type stars may not be as useful in detecting magnetic fields of planets orbiting inside the habitable zones of M-dwarf stars.  

\def\aj{{AJ}}                   
\def\araa{{ARA\&A}}             
\def\apj{{ApJ}}                 
\def\apjl{{ApJ}}                
\def\apjs{{ApJS}}               
\def\ao{{Appl.~Opt.}}           
\def\apss{{Ap\&SS}}             
\def\aap{{A\&A}}                
\def\aapr{{A\&A~Rev.}}          
\def\aaps{{A\&AS}}              
\def\azh{{AZh}}                 
\def\baas{{BAAS}}               
\def\jrasc{{JRASC}}             
\def\memras{{MmRAS}}            
\def\mnras{{MNRAS}}             
\def\pra{{Phys.~Rev.~A}}        
\def\prb{{Phys.~Rev.~B}}        
\def\prc{{Phys.~Rev.~C}}        
\def\prd{{Phys.~Rev.~D}}        
\def\pre{{Phys.~Rev.~E}}        
\def\prl{{Phys.~Rev.~Lett.}}    
\def\pasp{{PASP}}               
\def\pasj{{PASJ}}               
\def\qjras{{QJRAS}}             
\def\skytel{{S\&T}}             
\def\solphys{{Sol.~Phys.}}      
\def\sovast{{Soviet~Ast.}}      
\def\ssr{{Space~Sci.~Rev.}}     
\def\zap{{ZAp}}                 
\def\nat{{Nature}}              
\def\iaucirc{{IAU~Circ.}}       
\def\aplett{{Astrophys.~Lett.}} 
\def\apspr{{Astrophys.~Space~Phys.~Res.}}   
\def\bain{{Bull.~Astron.~Inst.~Netherlands}}    
\def\fcp{{Fund.~Cosmic~Phys.}}  
\def\gca{{Geochim.~Cosmochim.~Acta}}        
\def\grl{{Geophys.~Res.~Lett.}} 
\def\jcp{{J.~Chem.~Phys.}}      
\def\jgr{{J.~Geophys.~Res.}}    
\def\jqsrt{{J.~Quant.~Spec.~Radiat.~Transf.}}   
\def\memsai{{Mem.~Soc.~Astron.~Italiana}}   
\def\nphysa{{Nucl.~Phys.~A}}    
\def\physrep{{Phys.~Rep.}}      
\def\physscr{{Phys.~Scr}}       
\def\planss{{Planet.~Space~Sci.}}           
\def\procspie{{Proc.~SPIE}}     

\let\astap=\aap
\let\apjlett=\apjl
\let\apjsupp=\apjs
\let\applopt=\ao

\end{document}